\documentclass[prl,aps,twocolumn,superscriptaddress]{revtex4-2}

\usepackage{latexsym} \usepackage{graphicx}
\usepackage{amssymb} \usepackage{bm} \usepackage{framed}
\usepackage{lipsum} \usepackage{color} \usepackage{amsmath}
\usepackage{footmisc} \usepackage{hyperref}  \usepackage{url}
\usepackage{bm}

\newcommand{\vect}[1]{\bm{#1}}
\newcommand{\mc}[1]{{\cal #1}}

\begin{document} 

\title{A topological fluctuation theorem} 

\author{Beno\^{i}t Mahault}
\affiliation{Max Planck Institute for Dynamics and Self-Organization, 37077 G\"ottingen, Germany}

\author{Evelyn Tang}
\affiliation{Max Planck Institute for Dynamics and Self-Organization, 37077 G\"ottingen, Germany}

\author{Ramin Golestanian}
\email{ramin.golestanian@ds.mpg.de}
\affiliation{Max Planck Institute for Dynamics and Self-Organization, 37077 G\"ottingen, Germany}
\affiliation{Rudolf Peierls Centre for Theoretical Physics, University of Oxford, Oxford OX1 3PU, United Kingdom}

\begin{abstract}
Fluctuation theorems specify the non-zero probability to observe negative entropy production, contrary to a naive expectation from the second law of thermodynamics. For closed particle trajectories in a fluid, Stokes theorem can be used to give a geometric characterization of the entropy production. Building on this picture, we formulate a topological fluctuation theorem that depends only by the winding number around each vortex core and is insensitive to other aspects of the force. The probability is robust to local deformations of the particle trajectory, reminiscent of topologically protected modes in various classical and quantum systems. We demonstrate that entropy production is quantized in these strongly fluctuating systems, and it is controlled by a topological invariant.  
We demonstrate that the theorem holds even when the probability distributions are non-Gaussian functions of the generated heat.
\end{abstract}

\maketitle 

\section*{Introduction}

Concept from topology have played a key role in understanding a wide range of physical phenomena by providing an intuitive and mathematically rich effective description of the system \cite{Goldbart:2019wv,nakahara2003geometry}. In classical systems, topology has played a crucial role, for instance, in characterizing 2D turbulence \cite{Onsager1949} and defect-mediated phase transitions \cite{Kosterlitz_1972}. Topological defects abound in soft matter systems, such as dislocation and disclination pairs in 2D melting \cite{PhysRevB.19.2457} and various liquid crystalline systems \cite{ProstdeGennes1993}, and have recently featured in a variety of intrinsically nonequilbrium active matter systems \cite{UchidaPRL2010,ThampiPRL2013,ThampiEPL2014,PhysRevX.7.031039,shankar2021topological}. An early example of topological protection arose from use of the Gauss-Bonnet theorem in the physics of membranes \cite{refId0}. More recent work examined relevant topological invariants in mechanical lattices \cite{Kane2014} or dissipative systems in continuous space~\cite{Delplace1075} or lattice models with underlying periodic structure \cite{DasbiswasE9031,PhysRevLett.122.118001,sone2020exceptional,muru17,PhysRevE.93.042310,TangPRX2021}. Many such systems exhibit non-Hermitian properties such as exceptional points \cite{Heiss:2012dx,PhysRevLett.123.205502,You19767,PhysRevX.10.041009,FreyPRL2021} and a non-zero topological vorticity of the edge state \cite{Delplace1075,TangPRX2021}. The topological systems that support protected edge states are robust to disorder and perturbations, providing a key towards understanding such phenomena.  

Fluctuations can give rise to particle trajectories with negative entropy, which appears to contradict a fundamental law of macroscopic physics. Fluctuation theorems provide a quantitative, probabilistic prediction for this negative entropy production. These universal laws are valid even during processes that drive systems far from equilibrium. While the probability to observe such `violations' is typically exponentially small in the relevant system size, it can be appreciable in small systems. The first theorem was discovered over two decades ago by computer simulations and justified heuristically \cite{evans93}, then proven for a large class of systems \cite{Gallavotti95,Kurchan_1998}. This led to a class of relations dealing with the distribution functions of thermodynamic quantities such as exchanged heat, applied work or entropy production \cite{Jarzynski_rev}. Further work extended the concepts of thermodyamics to the level of individual trajectories \cite{10.1143/PTPS.130.17,PhysRevLett.95.040602}. Another key insight related entropy production in the medium to that part of the stochastic action which determines the weight of trajectories that is odd under time reversal \cite{Crooks1999PRE,Maes2003}. These and other developments opened the possibility for experimental or numerical measurements on the single molecule level, providing verification of the fluctuation theorems \cite{Hummer3658,Seifert_2012}. Notably, critical insights were obtained for the behavior of bio-molecules \cite{Ritort_intro}. Brownian dynamics of tracers in the presence of vortex-like singularities has been studied with an empasis on the winding number distribution and extensions to entanglement problems in polymer physics \cite{Spitzer1958,Edwards_1967,Prager1967,Drossel1996,Grosberg2003}. However, no connection has so far been made to stochastic thermodynamics.

\begin{figure*}[t!]
\includegraphics[width=.8\linewidth]{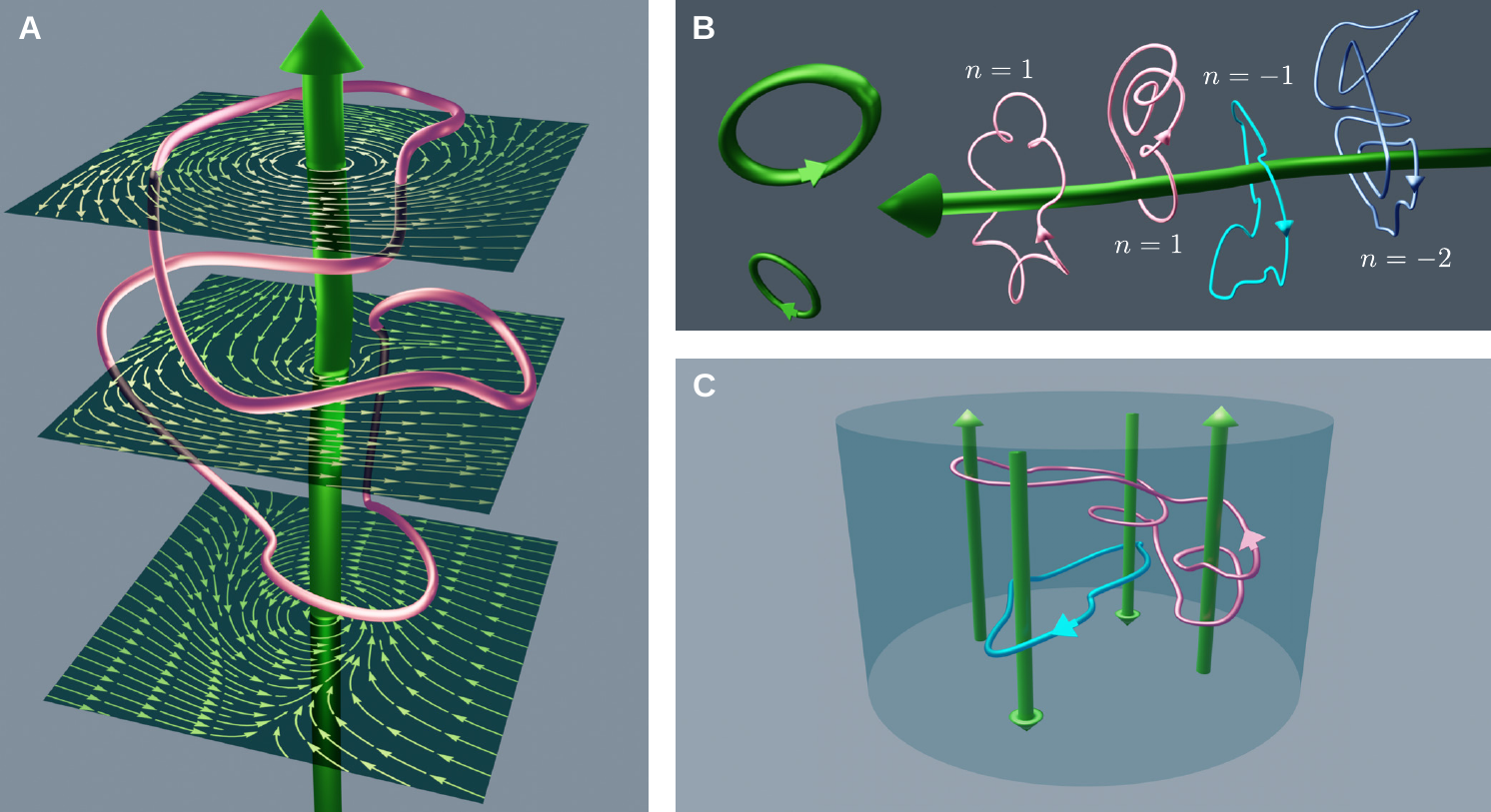}
\caption{(\textbf{A}) A tracer particle undergoes diffusion and drift in a force-field whose non-conservative component is generated by a vortex line (thick green), tracing out a stochastic trajectory (thin irregular loop). The force-field has the same rotational component throughout but varying irrotational components (the cross-sections depict force-field streamlines) -- however closed loops pick up only the rotational component. (\textbf{B}) 
Different particle trajectories around a single vortex line (green) can be characterized by the winding number $n$. The leftmost two are topologically equivalent with $n=1$, while the rightmost two have the opposite sign and winding numbers $n=-1$ and $n=-2$, respectively. (\textbf{C}) Force-fields can contain multiple vortex-cores (thick green lines or rings), around which particles can trace closed trajectories (thin irregular loops).} \label{fig:schematic}
\end{figure*}

Here, we identify a topological invariant that predicts observable quantities, in a strongly fluctuating system without underlying periodic structure. Our analysis allows us to quantify the ratio of particles with negative entropy production, purely as a function of winding number around vortex cores. We build on the geometrical properties of individual trajectories under the influence of external forces. This is studied in the context of a stochastic particle in a force-field, which is ubiquitous in nature (see Fig.~\ref{fig:schematic}). In particular, a particle moving around a closed path picks out only the non-conservative component of the force-field, which gives the entropy production. This allows us to formulate a topological fluctuation theorem based only on the vortex winding number. While previous work looked at entropy production in flow-fields \cite{PhysRevLett.100.178302,TangNJP2020}, the topological equivalence of particle trajectories and its consequences have not been studied. We begin with a general statement of the theorem using the examination of entropy production in the medium. 
We then demonstrate it in various examples including one or several vortices by calculating 
the corresponding exact winding number distribution or performing Brownian dynamics simulations. 
We find that even when the winding number distributions are non-Gaussian functions of the vortex circulation, the theorem holds exactly.

\section*{Results}

Consider the motion in $d$ dimensions of a tracer particle with diffusivity $D$ and mobility $\mu$ in a stationary force field $\vect{F}(\vect{r})$, where $D=\mu k_{\rm B} T$ with $k_{\rm B} T$ representing the thermal energy. The stochastic dynamics of the particle is characterized as a function of time $0 \leq \tau \leq t$ by its trajectory $\vect{r}(\tau)$, which satisfies the Langevin equation
\begin{equation}
\dot{\bm r}(\tau) =\mu \vect{F}(\vect{r}(\tau))+\sqrt{2 D}\, {\bm \xi}(\tau),\label{eq:Lang}
\end{equation}
where ${\bm \xi}$ represents a $d$-dimensional Gaussian white noise of zero mean and unit variance. Using the Helmholtz-Hodge decomposition, the force field can generally be separated into a conservative component that derives from a potential $U(\vect{r})$ and a rotational component $\vect{f}(\vect{r})$ that satisfies ${\bm \nabla}\cdot \vect{f}(\vect{r})=0$, i.e., $\vect{F}(\vect{r})=-{\bm \nabla} U(\vect{r})+\vect{f}(\vect{r})$. 
Let us assume that the rotational component of the force field is generated by a topological defect, such as the vortex line shown in Fig. \ref{fig:schematic}A for the physically relevant case $d = 3$.  
In the general case such a defect takes the form of codimension $2$ manifold $\Sigma$ giving rise to a vorticity field satisfying
\begin{equation} \label{eq:charge}
Q \equiv \int_S {\rm d} r^i \wedge  {\rm d} r^j \left(\partial_i f_j-\partial_j f_i\right) ,
\end{equation}
where $S$ is an arbitrary smooth surface intersecting $\Sigma$ transversally at a single point, the constant $Q$ is the strength of the vortex and summation is performed over $i<j$ (distinct pairs)~\cite{ShashikanthJMP2012,KHESIN2013135}.
For such a vortex-induced force-field, we derive a topological fluctuation theorem in terms of the probability $p(n,t)$ for any closed trajectory of duration $t$ to wind $n$ times around the vortex, which reads
\begin{equation}
\frac{p(-n,t)}{p(n,t)}=\exp\left(-\frac{n Q}{k_{\rm B} T}\right),
\label{eq:singlevor}
\end{equation} 
for any $U(\vect{r})$ (see Fig.~\ref{fig:schematic}A for a depiction of a representative force profile and closed loop in $d=3$). 
Here, $Q$ corresponds to the quantum of heat generated via the closed trajectories of the tracer particle that enclose the vortex domain. It is helpful to define $\gamma \equiv {Q}/{(k_{\rm B} T)}$ as the quantum of entropy production (in units of $k_{\rm B}$). Figure \ref{fig:schematic}B shows a number of exemplar closed trajectories with different winding numbers. Note that the theorem [Eq.~\eqref{eq:singlevor}] is valid at any time $t$. Therefore, while the positive and negative winding number distributions are expected to evolve with time and be affected by conservative contributions from the force, their ratio is topologically protected. 

\begin{figure*}[t]
\includegraphics[width=0.8\linewidth]{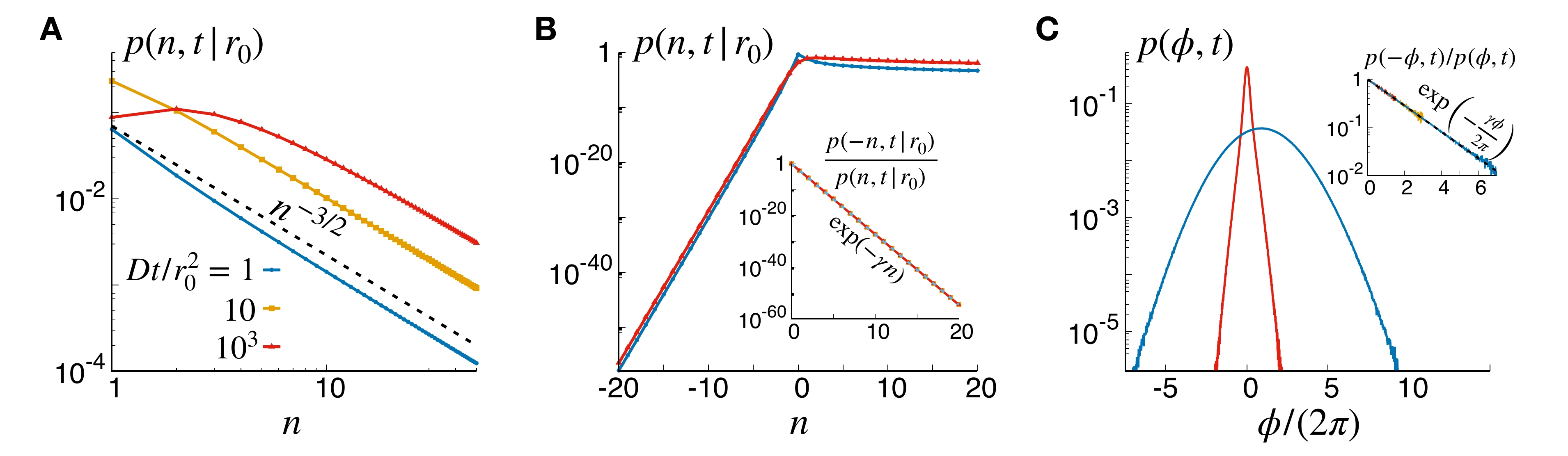}
\caption{The topological fluctuation theorem for a single vortex.
(\textbf{A}) The exact winding number distribution for a free vortex exhibits a power law scaling for large positive $n$ values.
(\textbf{B}) At all times, the ratio of negative to positive winding number distributions maps onto the theoretical 
prediction~\eqref{eq:singlevor} (inset) resulting in an exponential scaling of $p(n,t|r_0)$ for large negative windings.
(\textbf{C}) In the presence of boundaries restricting the particle motion inside a disk of radius $R$ and outside the vortex center, 
the winding angle distribution shows exponential tails at small times (red curve, $Dt/R^2 = 0.1$), while it becomes Gaussian for larger $t$ (blue curve, $Dt/R^2 = 10$). The theorem holds irrespectively of the shape of the distribution (inset, the orange curve corresponds to $Dt/R^2 = 1$).
In (\textbf{A},\textbf{B}) $\gamma=2\pi$, while in (\textbf{C}) $\gamma=0.1\times2\pi$.}
\label{fig:fp}
\end{figure*}

The derivation of the topological fluctuation theorem starts from the probability distribution of the particle to be at position $\vect{x}$ after time $t$ starting initially from $\vect{x}_0$, which can be be found via
     \begin{equation}
 \mc{P}(\vect{x},t|\vect{x}_0,0) = \int_{\vect{r}(0)=\vect{x}_0}^{\vect{r}(t)=\vect{x}}\mc{D}\vect{r}(\tau)P[\vect{r}(\tau)|\vect{x}_0],\label{eq:probfull}
     \end{equation}
where the probability of a specific stochastic trajectory is defined in terms of the Onsager-Machlup action associated with Eq. \eqref{eq:Lang} as
  \begin{align}
    P[\vect{r}(\tau)|\vect{x}_0]=& \mc{N}\exp\left(-\int^t_0{\rm d}\tau \left[ \frac{1}{4D}\Big(\vect{\dot{r}}(\tau) - \mu \vect{F}(\vect{r}(\tau))\Big)^2 \right. \right. \nonumber \\
   & \qquad\qquad\qquad\qquad \left.\left. +\frac{\mu}{2}\nabla\cdot \vect{F}(\vect{r}(\tau))\right]\right), 
     \label{eq:probtraj}
     \end{align}
where $\mc{N}$ is a normalization factor, and the Stratonovich convention is implied. 
Defining the kinematically-reversed, or backward, trajectory $\tilde{\vect{r}}(\tau)$ as $\tilde{\vect{r}}(\tau)\equiv\vect{r}(t-\tau)$,
we then find from Eq.~\eqref{eq:probtraj} that the ratio between the probabilities of the backward and forward paths is given by the part of the stochastic action that is odd under this transformation, namely,
\begin{equation}
\frac{P[\tilde{\vect{r}}(\tau)|\vect{x}]}{P[\vect{r}(\tau)|\vect{x}_0]} = \exp\left(-\Gamma\right), 
\label{eq:epmed}
\end{equation}
where $\Gamma \equiv \frac{1}{k_{\rm B} T} \int_0^t {\rm d}\tau \,\dot{\vect{r}}(\tau) \cdot \vect{F}(\vect{r}(\tau))$ corresponds to the heat generated (in units of $k_{\rm B} T$), or entropy production, during the motion of the particle along the trajectory.
We then consider a closed trajectory described by a curve $C$ (by setting $\vect{x}=\vect{x}_0$) which does not intersect with $\Sigma$, and implement the Stokes Theorem $\int_{\partial S} \omega= \int_{S} {\rm d} \omega$ for the 1-form 
$\omega=F_{i}{\rm d} r^i$ over the surface $S$ enclosed by $C$ \cite{nakahara2003geometry}, namely 
\begin{align}
\Gamma & =  \frac{1}{k_{\rm B} T} \oint_C {\rm d}\vect{r}\cdot \vect{F}\left(\vect{r}\right) =  \frac{1}{k_{\rm B} T} \int_{\partial S=C} {\rm d} r^i F_{i} \nonumber \\
& =  \frac{1}{k_{\rm B} T} \int_{S}  {\rm d} r^i \wedge  {\rm d} r^j \left(\partial_i F_j-\partial_j F_i\right).
\label{eq:circ}
\end{align}
It is manifest in Eq. \eqref{eq:circ} that $\Gamma$ picks out only the rotational component of the force.
Moreover, using the decomposition $\bm F = -\nabla U + \bm f$ and Eq.~\eqref{eq:charge} directly yields $\Gamma = \gamma n$,
with $n$ the number of times that the trajectory winds around the vortex singularity. We thus find that $\Gamma$ is independent of the initial position $\vect{x}_0$ or the specific shape of $C$ so long as it corresponds to the same winding number $n$. Hence, $\Gamma$ is identical for topologically equivalent curves (see Fig. \ref{fig:schematic}B), and independent of the conservative component of the force-field.

To complete the calculation, we can therefore integrate Eq.~\eqref{eq:epmed} over all initial positions $\{\vect{x}_0\}$ and topologically equivalent loops
to obtain a topological fluctuation theorem for vortex-induced force fields
\begin{equation}
\frac{p(-n,t)}{p(n,t)}=\exp\left(-{n\gamma}\right),
\label{eq:singlevor-2}
\end{equation} 
where $p(n,t) \propto \int {\rm d}\vect{x}_0 \pi(\vect{x}_0) \oint_{\vect{r}(0)=\vect{x}_0}\mc{D}\vect{r}(\tau) P[\vect{r}(\tau)|\vect{x}_0] \delta(\Gamma - \gamma n)$
is, up to a normalizing constant, the probability for any closed trajectory of length $t$ to wind $n$ times around the vortex axis,
while $\pi(\vect{x}_0)$ denotes the arbitrary distribution of $\vect{x}_0$. 

The above derivation is easily generalizable to the case of $m$ non-intersecting vortex branes. Denoting $\gamma_i$ and $n_i$ the 
dimensionless quantum of heat and winding number associated to a given vortex $i$,
the entropy production 
becomes $\Gamma=\sum_{i=1}^{m}\gamma_i n_i$.
Integrating Eq.~\eqref{eq:epmed} over initial conditions and trajectories which share the same set
of winding numbers  $\{n_i\}_{i=1,\ldots m}$,
the topological fluctuation theorem for multiple vortices then reads
\begin{equation}
\frac{p(\{-n_i\},t)}{p(\{n_i\},t)}=\exp\left(-{\sum_{i=1}^{m}\gamma_i n_i}\right),
\label{eq:multvor}
\end{equation}
where $p(\{n_i\},t)$ is the probability that a closed trajectory of length $t$ 
has wrapped $n_i$ times around each vortex $i$ for $i = 1, \ldots m$.

As for a single vortex, Eq.~\eqref{eq:multvor} is set by topology and thus holds at all times 
and for any conservative force affecting the particle motion.
Moreover, as both sides of Eq.~\eqref{eq:multvor} depend only on the set of winding numbers $\{n_i\}$ reached at time $t$,
the fluctuation theorem is insensitive to the history leading to $\{n_i\}$
 and in particular to the order in which the trajectory winds around each vortex.
Since any combination of winding numbers leading to the same value of $\Gamma$ leaves the r.h.s.\ of Eq.~\eqref{eq:multvor} unchanged, 
the latter can be further summed over all such configurations in order to get a weaker formulation of the theorem:
\begin{equation}
\frac{p(\Gamma = -\sum_{i=1}^{m}\gamma_i n_i,t)}{p(\Gamma = \sum_{i=1}^{m}\gamma_i n_i,t)}=\exp\left(-{\sum_{i=1}^{m}\gamma_i n_i}\right).
\label{eq:multvor_gamma}
\end{equation}

\paragraph*{Exact solution for a single vortex.} 
To shed some light on the topological fluctuation theorem~\eqref{eq:singlevor}, we now focus on the physically relevant case $d=3$
and consider the motion of a tracer particle in a flow field created by a single straight vortex line oriented along $\vect{e}_z$,
which is given in cylindrical coordinates ($r,\phi,z$) by
\begin{equation}
\vect{f}(\vect{r})=\frac{Q}{2\pi r}\,\vect{e}_{\phi},\qquad {\bm \nabla}\times\vect{f}(\vect{r})=Q \vect{e}_z\delta^2(\vect{r}_{\perp}),\label{eq:vortex}
\end{equation}
where $\vect{e}_{\phi} = (-\sin\phi,\cos\phi,0) $ and ${\vect{r}_\perp}=r(\cos\phi,\sin\phi)$,
so that from Eq.~\eqref{eq:circ} we can readily evaluate $\Gamma=\frac{1}{k_{\rm B} T} \int_S{\rm d} \vect{S} \cdot\nabla\times\vect{F}(\vect{r})= \gamma n$.
The simple case where the particle motion is restricted to a ring following the flow streamlines, such that the drive is effectively uniform, has been treated previously~\cite{PhysRevLett.95.040602}. This problem (generalizable to higher dimensions) corresponds to a biased random walk and the fluctuation theorem can be shown to result from the Gaussian form of the winding distribution. 

However, from the above derivation the relation~\eqref{eq:singlevor} holds even when the Brownian particle is allowed to move transversally to flow streamlines. Considering closed trajectories winding around a vortex line in free space, the Fokker-Planck equation describing the evolution of the distribution $\mc{P}(r,\phi,z,t)$ can be solved exactly~\cite{doi:10.1098/rsta.2018.0347},  such that the winding distribution is given, up to a normalizing constant, by (details in the Methods section below)
\begin{align}
p(n,t|r_0) & \propto \int_0^{\infty}{\rm d}u \, \textrm{Re}\left[e^{2i\pi nu}I_{k_u}\left(\frac{r_0^2}{2Dt}\right)\right] , \nonumber \\
k_u & = \sqrt{u^2+\frac{iu\gamma}{2\pi}},
\label{eq:exact_winding}
\end{align}
where $r_0$ denotes the initial radial position of the particle and $I_{\nu}$ is the modified Bessel function of the first kind, of order $\nu$.

A detailed examination of this probability distribution reveals that it is strongly non-Gaussian. Indeed, for positive winding numbers $p(n,t|r_0)$ is asymptotically scale free: $p(n,t|r_0) \underset{n\to+\infty}{\sim} n^{-3/2}$~\cite{doi:10.1098/rsta.2018.0347} (see Fig.~\ref{fig:fp}A), while, as a consequence of the fluctuation theorem~\eqref{eq:singlevor}, it decays exponentially as $n \to -\infty$.
This difference in scaling behaviors results in a strong asymmetry of the overall distribution between the positive and negative winding number sectors (see Fig.~\ref{fig:fp}B).  

\begin{figure}[t]
\includegraphics[width=\linewidth]{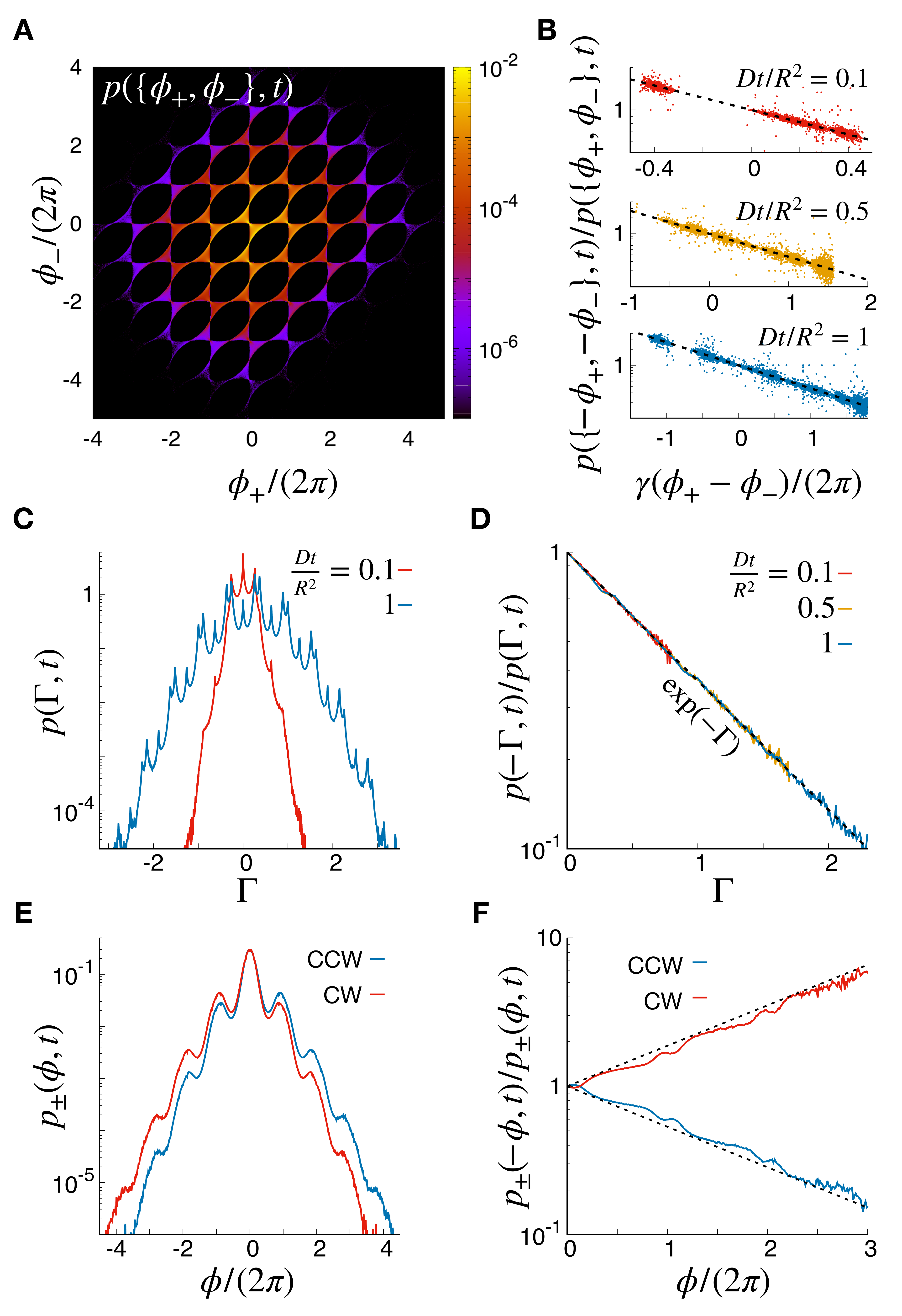}
\caption{The topological fluctuation theorem in the presence of multiple vortices. (\textbf{A},\textbf{B}) The strong formulation of the theorem is assessed by measuring the joint CCW and CW winding angle distribution $p(\{\phi_+,\phi_-\},t)$ (\textbf{A}) whose behavior satisfies Eq.~\eqref{eq:multvor} (\textbf{B}). (\textbf{C},\textbf{D}) The total entropy production distribution exhibits a complex structure (\textbf{C}), but obeys the theorem nonetheless (\textbf{D}). (\textbf{E},\textbf{F}) On the contrary, individual CCW(+) and CW(-) winding distributions (\textbf{E}) generally do not satisfy the theorem (\textbf{F}). In (\textbf{B},\textbf{D},\textbf{F}) the dashed black lines indicate the theoretical exponential scaling for the probability ratios. In (\textbf{A},\textbf{E},\textbf{F}) $Dt/R^2 = 1$ and in all plots we used $\gamma=0.1\times 2\pi$.} \label{fig:simus}
\end{figure}

\paragraph*{Computational verification of the theorem.}
For more complicated force profiles, solving the Fokker-Planck equation may not be possible, and one needs to resort to Brownian dynamics simulations.  
As the data shown in Fig.~\ref{fig:fp}B suggests, however, a numerical verification of the theorem is generally a demanding task due to the need for excessive sampling. One can nevertheless consider specific setups which allow for easier accumulation of winding number statistics.
For example, when the flow is effectively two-dimensional, such that all vortex lines are aligned with the third spatial direction, the theorem of Eq.~\eqref{eq:multvor} can be generalized to open trajectories and therefore to winding angles that are not multiples of $2\pi$ (see Methods).
Moreover, a potential $U(r)$ confining the trajectories inside a circular domain of finite radius $R$ leads to a faster convergence. Similarly, it will be beneficial to prevent the particles from reaching the vortex centers, where the winding angle is ill-defined; this will not affect the winding number probability ratios.

Let us consider a single vortex located at $r = 0$ and denote $\phi$ as the total winding angle spanned by the trajectory in a time $t$. For such a configuration, we find that for $t \ll R^2/D$ the distribution $p(\phi,t)$ exhibits exponential tails, while for $t \gg R^2/D$ the effect of confining boundaries leads to $p(\phi,t)$ being Gaussian (see Fig~\ref{fig:fp}C). As expected, the theorem is found to hold at all times, independently of the particular shape of the distribution. The less straightforward multi-vortex case is addressed considering a configuration of four counter rotating vortices of strengths $\pm Q$ all located at $r = R/2$ and $90$ degrees from each other, as depicted in Fig.~\ref{fig:schematic}D~\footnote{Except that for simplicity the particle motion in simulations is restricted to 2 dimensions.}.

To be able to sample winding number distributions with sufficiently accurate statistics, we define the joint probability $p(\{\phi_+,\phi_-\},t)$ of total winding angles $\phi_+$ and $\phi_-$ around respectively counter-clockwise (CCW) and clockwise (CW) rotating vortices. The distribution, shown in Fig.~\ref{fig:simus}A, exhibits a nontrivial, time-invariant, $2\pi$-periodic structure. In particular, while $p(\{\phi_+,\phi_-\},t)$ shows local maxima when both $\phi_+$ and $\phi_-$ are integer factors of $2\pi$, it vanishes up to numerical precision when both of them are odd integer factors of $\pi$. Despite this complex behavior, representing the ratio $p(\{-\phi_+,-\phi_-\},t)/p(\{\phi_+,\phi_-\},t)$ as a function of $\gamma (\phi_+ - \phi_-)$ reveals a clear exponential scaling at all accessible times (see Fig.~\ref{fig:simus}B), in agreement with Eq.~\eqref{eq:multvor}.

We now examine the total entropy production distribution $p(\Gamma,t)$. Figure~\ref{fig:simus}C shows that although it exhibits a Gaussian envelope, it features prominent modulations that can not be described in terms of simple functions. Nevertheless, the ratio $p(-\Gamma,t)/p(\Gamma,t)$ always verifies Eq.~\eqref{eq:multvor_gamma} at all times and for all winding angle configurations (see Fig.~\ref{fig:simus}D).

Lastly, our Brownian dynamics simulations allow us to record the individual winding distributions $p_\pm(\phi,t)$ associated with CCW(+) and CW(-) rotating vortices. As shown in Fig.~\ref{fig:simus}E, the two distributions show smooth damped oscillations and are symmetric with respect to each other. Examining the ratios $p_\pm(-\phi,t)/p_\pm(\phi,t)$, we find that they do {\it not} satisfy the theorem. Namely, we find in both cases that the probability ratio of winding oppositely to the direction set by the vortex flow to that of winding along it is always larger than predicted by the theorem (see Fig.~\ref{fig:simus}F). This striking feature is a direct consequence of the fact that in presence of many vortices, trajectories winding with an angle $\phi$ around a given vortex are not all topologically equivalent.

\section*{Discussion} 

We have demonstrated a topological fluctuation theorem that identifies a relevant topological invariant able to predict observable quantities, in a strongly fluctuating system without underlying periodic structure. The ratio of particle trajectories going against the flow to those going along it
is purely a function of heat generated along the vortices and is topologically protected against deformations of these trajectories. Thanks to this property, the theorem holds for any finite observation time and is insensitive to conservative contributions to the force-field such as a confining potential, which makes it generic and observable in realistic experimental conditions.

In the context of micro-machines, where some degrees of freedom are driven out-of-equilibrium by external forces and torques that are even under time-reversal, the fluctuation theorem constrains the probability ratio of negative and positive entropy production created over cycles performed by the machine. Remarkably, the theorem predicts that this ratio is topologically protected for closed cycles in phase space, as discussed recently for the one dimensional case in Ref.~\cite{box2020rotor}, thus providing some insight for the optimization and design of micro-machines.

Finally, it will be interesting to investigate how the above results generalize with the introduction of nonequilibrium activity, 
be it as a correlated bath or as persistence in the particle motion~\cite{PhysRevX.9.021009}. 

\acknowledgments
We thank Viktoryia Novak for assistance with graphic design. 

\bibliography{scibib}

\pagebreak

\onecolumngrid
\appendix

\setcounter{equation}{0}
\setcounter{figure}{0}
\renewcommand{\theequation}{S\arabic{equation}}
\renewcommand{\thefigure}{S\arabic{figure}}

\section*{Methods}

\subsection*{The exact winding number distribution for a single vortex.}

Here we provide the derivation of the exact winding number distribution in the case where the force-field is generated by a single vortex line.
The Fokker-Planck equation associated to the dynamics of a stochastic particle in the vortex field defined in Eq. \eqref{eq:vortex} is given by
\begin{equation}
\partial_t  \mc{P}(r,\phi,z,t)+\frac{\gamma D}{2\pi r^2}\partial_{\phi} \mc{P}(r,\phi,z,t)=D\left[\partial^2_{rr}+\frac{1}{r}\partial_r+\frac{1}{r^2}\partial^2_{\phi\phi} + \partial^2_{zz}\right] \mc{P}(r,\phi,z,t),\label{eq:fpploar}
\end{equation}
where at $t=0$ the distribution $\mc{P}(r,\phi,z,t)$ satisfies $ \mc{P}(r,\phi,z,0)=r^{-1}\delta(r-r_0)\delta(\phi - \phi_0)\delta(z - z_0)$
with $r_0 > 0$, while $\phi_0$ and $z_0$ are set to $0$ without loss of generality.
Due to the symmetries of the problem, solutions of the Fokker-Planck equation [Eq.~\eqref{eq:fpploar}] are written as
\begin{equation}
 \mc{P}(r,\phi,z,t) = \mc{P}_{\rm 2D}(r,\phi,t) \times \frac{1}{\sqrt{4\pi Dt}}\exp\left(-\frac{z^2}{4Dt}\right)\,,
\end{equation}
where $\mc{P}_{\rm 2D}(r,\phi,t)$ can itself be decomposed into separable functions of the form 
$e^{iu\phi} e^{-D \lambda^2 t} \rho(r)$ \cite{doi:10.1098/rsta.2018.0347}. 
Meanwhile, $\rho(r)$ satisfies
\begin{equation}
\rho''(r) + \frac{1}{r} \rho'(r) + \left(\lambda^2 - \frac{k_u^2}{r^2} \right)\rho(r) = 0 \; \text{ with } \; k_u^2 = u^2 + \frac{iu\gamma}{2\pi} \,.
\label{eq_rho}
\end{equation}
For what follows, we shall consider solutions for which ${\rm Re}\left( k_u \right) \ge 0$.
Solutions of Eq. \eqref{eq_rho} take the general form
\begin{equation}
\rho(r) = C_J(\lambda,u) J_{k_u}\left(\lambda r\right) + C_Y(\lambda,u) Y_{k_u}\left(\lambda r\right) \,,
\end{equation}
where $J_\nu$ and $Y_\nu$ are Bessel functions of the first and second kind respectively, of order $\nu$. 

Moreover, since for a vortex-generated field in open space the Brownian particle avoids the origin with probability 1~\cite{Spitzer1958},
the distribution satisfies at all times, angles and $z$, $\mc{P}(0,\phi,z,t)= \mc{P}(r\to +\infty,\phi,z,t)=0$. 
From this constraint, $C_Y$ can be set to $0$ in what follows.
The distribution then becomes
\begin{equation}
 \mc{P}_{\rm 2D}(r,\phi,t) = \int_{-\infty}^\infty {\rm d}u \, \int_0^\infty {\rm d}\lambda \, C_J(\lambda,u) J_{k_u}\left(\lambda r\right) e^{i u \phi} e^{-D \lambda^2 t}\,.
\end{equation}
In order to satisfy the initial condition $ \mc{P}_{\rm 2D}(r,\phi,0) = r^{-1} \delta(r - r_0) \delta(\phi)$, 
we use a closure relation for Bessel functions $\int_0^\infty {\rm d} x \, x J_\nu(x s) J_{\nu}(x v) = s^{-1} \delta(s-v)$, to obtain
\begin{equation}
 \mc{P}_{\rm 2D}(r,\phi,t) = \frac{1}{2\pi}\int_{-\infty}^\infty {\rm d}u \, \int_0^\infty {\rm d}\lambda \, \lambda J_{k_u}\left({\lambda r_0}\right) J_{k_u}\left({\lambda r}\right) e^{i u \phi} e^{-D \lambda^2 t}\,.
\label{eq_Pgen}
\end{equation}

To simplify this further, we use the following relation \cite{gradshteyn2007}
\begin{equation}
\int_0^{\infty} {\rm d}t \, t J_\nu(\alpha t) J_\nu(\beta t) e^{-p^2 t^2} = \frac{1}{2 p^2} e^{-\frac{\alpha^2 + \beta^2}{4p^2}} I_\nu\left( \frac{\alpha \beta}{2p^2} \right) \quad (\alpha, \beta > 0,\; \textrm{Re}(\nu)> -1) \,,
\end{equation}
where $I_\nu$ is the modified Bessel function of the first kind, of order $\nu$. 
This gives us
\begin{equation}
 \mc{P}_{\rm 2D}(r,\phi,t)=\frac{e^{-\frac{r^2+r_0^2}{4Dt}}}{4\pi Dt}\int_{-\infty}^{\infty}{\rm d}u\, e^{iu\phi}I_{k_u}\left(\frac{r_0r}{2Dt}\right)\,,\label{eq:fullfp}
\end{equation}
where $I_{\nu}$ is the modified Bessel function of the first kind, of order $\nu$.

For closed trajectories, the winding number distribution associated with a particular initial position $(r_0,\phi_0=0,z_0=0)$ is 
$p(n,t|r_0) \equiv \tfrac{1}{N}\mc{P}(r_0,\phi=2\pi n,0,t)$
with $N\equiv \sum_{n=-\infty}^{+\infty} p(n,t|r_0)$, giving Eq.~\eqref{eq:exact_winding}.

\subsection*{Brownian dynamics simulations.}

Here, we provide the details of the Langevin dynamics simulation method that we have used. Denoting ${\bm r}(\tau)$ the particle position at time $\tau$, which is discretized in units of ${\rm d}\tau$, it is updated by means of an Euler-Maruyama scheme:
\begin{equation}
{\bm r}(\tau + {\rm d}\tau) - {\bm r}(\tau) = \mu \left[{\bm f}({\bm r}(\tau) ) - {\bm \nabla} U({\bm r}(\tau)) \right]{\rm d}\tau + \sqrt{2D {\rm d}\tau} \, {\bm \xi}(\tau) \,,
\label{eq:Langevin}
\end{equation}
where ${\bm f}({\bm r})$ denotes the applied vortex field, $\mu$ is the mobility of the particle, $U$ is a confining potential, and ${\bm \xi}$ is a Gaussian white noise vector with unit variance.

For simplicity, the simulations were restricted to 2 dimensions. In this particular case, considering $m$ vortices the flow field ${\bm f}({\bm r})$ takes the general form
\begin{equation}
\frac{\mu {\bm f}({\bm r}) }{k_{\rm B} T}= \sum_{i=1}^m \frac{\gamma_i}{2\pi} {\bm \nabla} \phi_i  \qquad \phi_i \equiv {\rm arg}\left({\bm r} - {\bm r}_i \right) \,,
\end{equation}
where ${\bm r}_i$ and $\gamma_i$ are respectively the position and effective (dimensionless) strength of each vortex.
Therefore, assuming that a given trajectory $C$ does not visit the vortex centers ${\bm r}_i$, we can write
\begin{equation}
\frac{1}{k_{\rm B} T} \int_C {\rm d}{\bm r} \cdot {\bm F}({\bm r})=\frac{1}{k_{\rm B} T}  \int_C {\rm d}{\bm r} \cdot \left[{\bm f}({\bm r}) - {\bm  \nabla} U\right] =  \sum_{i=1}^m \frac{\gamma_i}{2\pi} \Delta \phi_i - \frac{\Delta U}{k_{\rm B} T}  \,,
\label{eq:supp_gamma}
\end{equation}
where $\Delta \phi_i$ and $\Delta U$ denote the differences in angles and energy between the final and initial
points of $C$.

\begin{figure}[t]
\includegraphics[width=0.8\linewidth]{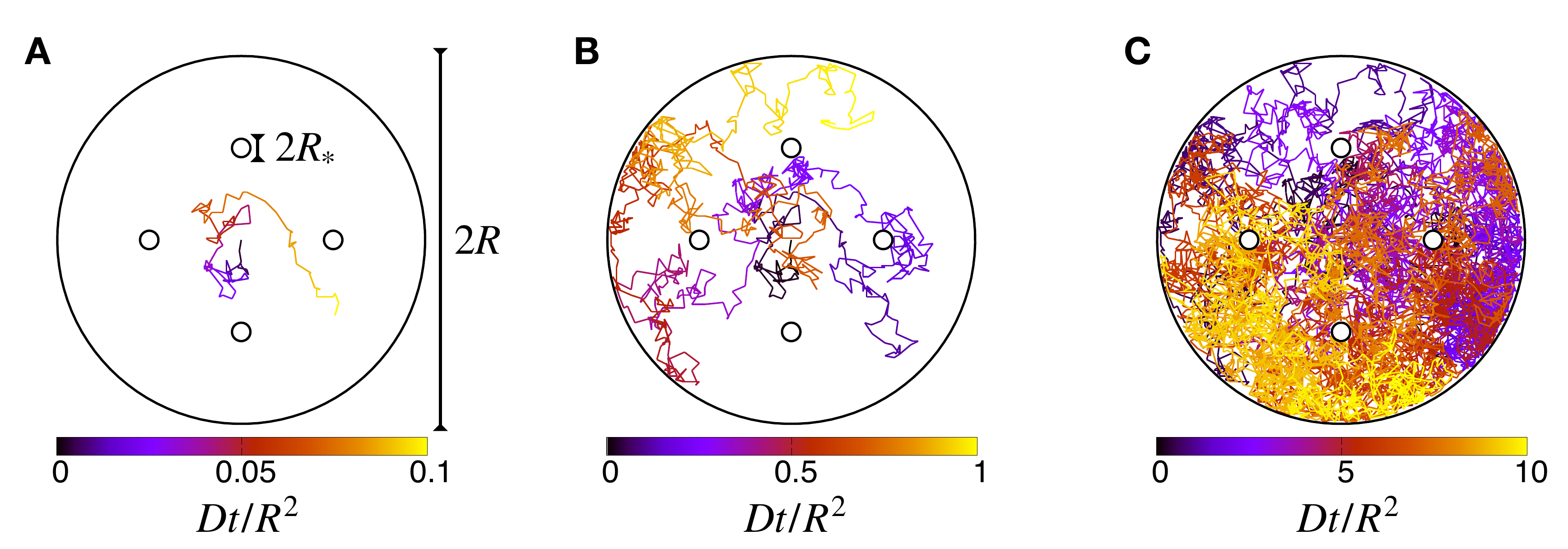}
\caption{Snapshots of a simulated trajectory starting at $r=0$ with four counter rotating vortices, with the color labeling the rescaled time.
When $t \ll R^2/D$ the particle has only explored a small portion of the available domain ({\bf A}), such that it encounters the external boundaries only at larger times ({\bf B}), while for $t \gg R^2/D$ its trajectory covers all the disk of radius $R$ ({\bf C}). In all panels the black lines mark the boundaries set by the confining potential.} 
\label{fig:trajs}
\end{figure}

The potential $U$ was chosen to ensure that all trajectories are confined inside a disk of radius $R$,
and that they cannot reach the vortex centers where the winding angles are not defined.
Namely, we used the following form for the potential
\begin{equation}
U({\bm r}) = \frac{k}{2} \times \left\{ \begin{array}{ccl} 
(|{\bm r}| - R)^2 & & |{\bm r}| \ge R \\
( |{\bm r}-{\bm r}_i| - R_*)^2 & & |{\bm r}-{\bm r}_i| \le R_* \\
0 & & {\rm otherwise}\end{array}\right. \,.
\end{equation}
When the length-scale $\ell \equiv \sqrt{k_{\rm B} T / k}$ is small compared to $R_*$ and $R$, 
the trajectories do not penetrate the limiting regions with $r > R$ and $|{\bm r}-{\bm r}_i| < R_*$ (see Fig.~\ref{fig:trajs}), 
such that $\Delta U$ in Eq.~\eqref{eq:supp_gamma} can be set to $0$.

From the derivation given before, it is thus clear that
\begin{equation}
\frac{p(\{-\phi_i\},t)}{p(\{\phi_i\},t)} = \exp\left( -\sum_{i=1}^m \frac{\gamma_i}{2\pi} \Delta \phi_i \right),
\end{equation}
such that for this special simulation setup the theorem holds for open trajectories as well.

Rescaling space and time in Eq.~\eqref{eq:Langevin}, we set $D = 1$ and $\mu k = 10^3$. With these units, we used in all simulations $\gamma = 0.1\times 2\pi$, ${\rm d}t/(\mu k) = 10^{-5}$, $R/\ell = 10^{3}\sqrt{10}$, and $R_*/\ell = 50\sqrt{10}$. We have verified that varying moderately these parameters did not affect our results. Both the data shown in Figs.~\ref{fig:fp} and~\ref{fig:simus} were obtained by sampling the winding distributions over $10^8$ independent trajectories initialized uniformly in the region of space where $U=0$.

\end{document}